\documentclass[aps,prl,reprint,amsmath,amssymb]{revtex4-2}
\usepackage{graphicx}
\usepackage{hyperref}
\usepackage{xcolor}
\hypersetup{colorlinks,allcolors=blue,bookmarksnumbered}
\usepackage{orcidlink}
\definecolor{Rpcorrect1}{rgb}{0, 0.5, 0}
\definecolor{Rpcorrect}{rgb}{1,0,0}

\begin{document}
\title{Tuning Density and Spin Ordering of Degenerate Fermi Gases in an Optical Cavity}
\author{Wei Qin}
\author{Yuan-Hong Chen\,\orcidlink{0009-0000-1273-5942}}
\author{Renyuan Liao\,\orcidlink{0000-0001-7412-3538}}
\email[Corresponding author:~]{ryliao@fjnu.edu.cn}
\affiliation{College of Physics and Energy, Fujian Normal University, Fujian Provincial Key Laboratory of Quantum Manipulation and New Energy Materials, Fuzhou 350117, China}
\affiliation{Fujian Provincial Engineering Technology Research Center of Solar Energy Conversion and Energy Storage, Fuzhou 350117, China}
\date{\today}

\begin{abstract}
We investigate a spin-degenerate Fermi gas coupled to a high-finesse optical cavity, where the competition between scalar and vectorial couplings is controlled by the relative polarization angle of the pump and cavity fields. We find that the phase transition threshold is synergistically determined by the scalar-vectorial coupling weight and Pauli blocking, with the latter dictating the critical pump lattice depth required for the onset of superradiance. For a two-component Fermi gas with opposite spins, the population ratio drives two distinct types of phase transitions corresponding to real-space phase separation: continuous and discontinuous. Nevertheless, the boundary of the phase transition remains fundamentally governed by the scalar-vectorial coupling competition. We clarify the impact of the relative polarization angle on phase transitions of the system; these results also apply to bosonic systems. Our results provide valuable theoretical insights for future experimental realizations.

\end{abstract}

\maketitle

\textit{Introduction.}\textemdash The coupling of ultracold quantum gases to optical cavities provides a powerful platform for studying many-body physics~\cite{Bloch2008, Ritsch2013, Farokh2021}. A prominent example is the superradiant self-organization quantum phase transition---realized in both Bose-Einstein condensates (BEC) and degenerate Fermi gases---which has been extensively investigated both experimentally and theoretically~\cite{Baumann2010, BENJ2018, STAM23, QinWei2024, ChenYu2015, ZhangXiaotian2021}. In such setups, effective long-range interactions between atoms are mediated by two-photon scattering processes between the cavity and the transverse pump fields, dominating the many-body phases~\cite{RMP2023}. Accordingly, extending such hybrid cavity-atom systems offers profound insights into the critical behavior of open quantum systems~\cite{Diehl2008, Chitra2015, Soriente2018}, nonequilibrium dynamics~\cite{Piazza14, Zwettler2025}, multicomponent interplay~\cite{Kroeze2018, Carl2023, LINQIN2026}, and thermodynamic phase transitions~\cite{ZhangYuanWei2013, Nairn2025}.

While extensive research has focused on coupling the external degrees of freedom of ultracold gases to optical cavities via atomic scalar polarizability~\cite{Leonard2017, Zupancic2019, LiXiangLiang2021, ChenShi2024}, introducing internal spin degrees of freedom in the presence of an external field allows the light-matter interaction to be decomposed into distinct scalar and vectorial components~\cite{LeKien2013}. Although the superradiant phase transition has been widely explored in spinor BEC~\cite{ZhangZhiqiang2018, Nishant2019, Morales2019, FanJingtao2020}, its counterpart in spin-dependent degenerate Fermi gases remains largely elusive~\cite{PanJianSong2015, FanJingtao2018, FengYanlin2019, FengYanlin2025}. In this Letter, we investigate the fundamental scenario of tuning the pump-field polarization to control the competition between these scalar and vectorial couplings, revealing the nontrivial features of the superradiant phase transition in spin-dependent fermions.  Unlike in bosonic systems, Pauli blocking dictates that the atomic response to the cavity field is collectively governed by the entire Fermi sea.

The central objective of our work is to unveil how the competition between scalar and vectorial couplings reshapes self-organized phase transitions in the presence of Fermi statistics. The paper is organized as follows. We begin by introducing our theoretical model and deriving the phase-transition conditions via mean-field theory. Our analytical results reveal that the pump polarization tunes the relative contributions of the scalar and vectorial couplings, while Fermi statistics dictates the susceptibility required to trigger the onset of superradiance. Both factors jointly set the phase transition threshold. Further, we extend the framework to a two-component Fermi gas with opposite spins. Steady-state numerical analysis demonstrates that the boundary for real-space phase separation is strictly anchored by the weight of the scalar and vectorial atomic polarizabilities , whereas the order of the phase transition is governed by the population ratio of two components.

\begin{figure}[t]
   \includegraphics[width=1\linewidth]{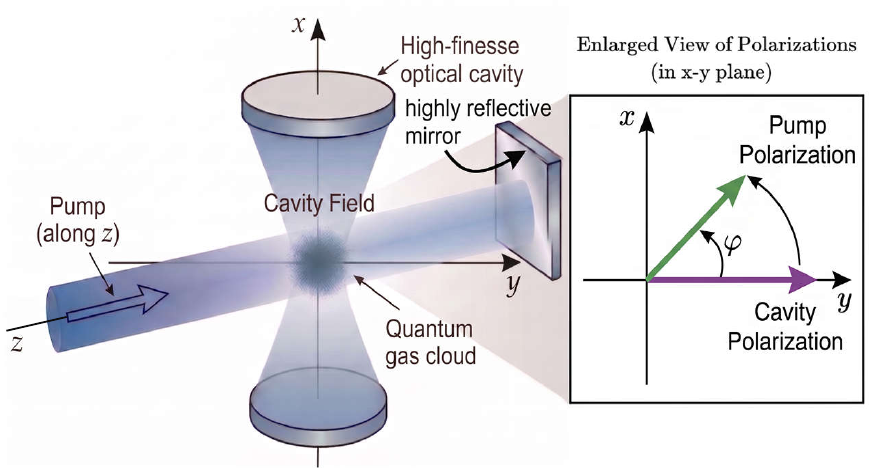}
   \caption{Schematic of the experimental setup. The degenerate Fermi gas is positioned at the center of the cavity axis. The high-finesse optical cavity is oriented along $\hat{x}$, and the pump laser is directed along $\hat{z}$. The cavity polarization (purple arrow) is taken to be along $\hat{y}$, and the pump polarization (green arrow) makes an angle $\varphi$ with respect to this direction.}
   \label{fig1}
\end{figure}

\textit{Model.}\textemdash  We consider spin-dependent degenerate Fermi gases confined within a high-fineness optical cavity, and driven by a transverse pump laser. As shown in Fig.~\ref{fig1}, the cavity field is along the $x$-axis and its polarization is in the $x$-axis direction. As the polarization of transverse pump is adjustable, we define $\varphi$ as a relative polarization angle between the pump and the cavity fields. Consequently, in above circumstance the atom-light interaction contains both scalar and vectorial parts~\cite{LeKien2013, Morales2019}. To proceed, we adiabatically eliminate the excited states of the atoms because the pump frequency $\omega_p$ is far detuned from the atomic transition frequency $\omega_a$~\cite{Bhaseen2012}, resulting in an effective Hamiltonian~\cite{Landini2018,supp}: \nocite{Sieberer2016,Kamenev2023}%
\begin{equation}
\hat{H}_{\mathrm{eff}}=-\Delta_c\hat{a}^\dagger\hat{a}+\int\mathrm{d}\mathbf{r}\Psi_{m_F}^\dagger(\mathbf{r}) \big ( \hat{H}_0+\hat{H}_{I,m_F} \big ) \Psi_{m_F}(\mathbf{r}),
\label{eq:single_component_H}
\end{equation}
with
\begin{equation}
    \hat{H}_0=\frac{\hat{\mathbf{p}}^2}{2m}-\mathrm{V}(\mathbf{r}),
\end{equation}
\begin{equation}
    \hat{H}_{I,m_F}=-\mathrm{U}(\mathbf{r})\hat{a}^\dagger\hat{a}-\eta(\mathbf{r})\lambda_{m_F}\big(e^{i\phi_{m_F}}\hat{a}^\dagger+e^{-i\phi_{m_F}}\hat{a}\big).
\end{equation}
In the above, $\hat{a}^\dagger$ ($\hat{a}$) is the creation (annihilation) operators of photons in cavity mode, and $\Psi_{m_F}^\dagger$ ($\Psi_{m_F}$) is the creation (annihilation) operators of fermionic atoms with mass $m$, momentum operator $\hat{\mathbf{p}}=-i\hbar\nabla$ and magnetic sublevels $m_F$. $\Delta_c=\omega_p-\omega_c$ denotes cavity-pump detuning with the cavity frequency $\omega_c$. We have defined the pump potential $\mathrm{V}(\mathbf{r})=\mathrm{V}_0\cos^2(\mathbf{k}_p\cdot\mathbf{r})$, the cavity field $\mathrm{U}(\mathbf{r})=\mathrm{U}_0\cos^2(\mathbf{k}_c\cdot\mathbf{r})$, and the coefficient of pump-cavity interference $\eta(\mathbf{r})=\eta_0\cos(\mathbf{k}_p\cdot\mathbf{r})\cos(\mathbf{k}_c\cdot\mathbf{r})$ with $\eta_0=\sqrt{\mathrm{V}_0\mathrm{U}_0}$. The information of $\varphi$ is encoded in the symbols $\lambda_{m_F}$ and $\phi_{m_F}$: $\lambda_{m_F}=\sqrt{\cos^2\varphi+(\alpha_v m_F \sin\varphi/2\alpha_s F)^2}$ and $\phi_{m_F}=\mathrm{arctan}(\alpha_v m_F \tan\varphi/2\alpha_s F)$~\cite{supp}. Here $F=3/2$ is the total angular momentum manifold stemming from the $D_2$ line of a $^{6}\mathrm{Li}$ atom, so $m_F=\{\pm3/2, \pm1/2\}$ is chosen~\cite{Zwettler2025-2}. $\alpha_s$ and $\alpha_v$ are, respectively, the scalar and vectorial atomic polarizabilities, and we fix the value $\alpha_v/\alpha_s=0.928$. The wave vector of the pump beam is given by $\mathbf{k}_p=k_0\hat{e}_z$ and that of the cavity field is given by $\mathbf{k}_c=k_0\hat{e}_x$. For simplicity, we consider the weak-coupling limit, set $\hbar=1$ and choose the recoil energy $E_R=\hbar^2k_0^2/2m$ as the basic energy scale.

\textit{Method.}\textemdash The dynamics of the photon operator $\hat{a}$ is governed by the Lindblad master equation. Within the mean-field approach~\cite{Manzano2020}, $\hat{a}$ is replaced by its expectation value $\alpha=\langle \hat{a} \rangle$. We seek a steady state by the condition $\partial_t\alpha=0$ and obtain
\begin{equation}
 \alpha \approx -\frac{\eta_0\lambda_{m_F}}{\tilde{\Delta}_c}e^{i\phi_{m_F}}\Theta_{m_F}.
 \label{eq:ave_pho_num}
\end{equation}
The cavity losses $\kappa$ has been neglected in present case of the high-finesse cavity, and key features of our result survive  in the weak dissipative limit~\cite{Baumann2010,Baumann2011}. We have defined the density order parameter $\Theta_{m_F}=\int\mathrm{d}\mathbf{r}n_{m_F}(\mathbf{r})\eta(\mathbf{r})  /\eta_0$ and the shifted detuning $\tilde{\Delta}_c=\Delta_c+\int\mathrm{d}\mathbf{r}\mathrm{U}(\mathbf{r})n_{m_F}(\mathbf{r})$, with $n_{m_F}(\mathbf{r})=\langle \Psi_{m_F}^\dagger(\mathbf{r})\Psi_{m_F}(\mathbf{r}) \rangle$.

The partition function within the Euclidean functional integral formalism is $\mathcal{Z}=\int \mathcal{D}[\bar{\psi}_{m_F},\psi_{m_F}]e^{-S}$, with the action $S=\int_0^\beta\mathrm{d}\tau\int\mathrm{d}\mathbf{r}\big [\bar{\psi}_{m_F}(\partial_\tau-\mu_{m_F}) \psi_{m_F}+H_{\mathrm{eff}}(\bar{\psi}_{m_F},\psi_{m_F})  \big]$ adopted from Refs.~\cite{Stoof2009, Altland2023}. Here, $\psi_{m_F}$ ($\bar{\psi}_{m_F}$) is Grassmann variables representing the fermionic modes, $\beta=1/k_BT$ is the inverse temperate and $\mu_{m_F}$ is the chemical potential. According to the Landau theory, we calculate the free energy of the system $F=-\ln\mathcal{Z}/\beta+\mu_{m_F}N_{m_F}$ around $\alpha=0$, to determine the continuous transition between the normal phase and the superradiant phase. Meanwhile, expanding it up to the quadratic order in $\Theta_{m_F}$,  we obtain~\cite{supp}
\begin{equation}
F=F_0-\frac{\eta^2_0\lambda^2_{m_F}}{\tilde{\Delta}_c}(1+\chi_{m_F}\lambda^2_{m_F}\frac{4}{\tilde{\Delta}_c} )\Theta_{m_F}^2,
\label{eq:GrandPotential}
\end{equation}
where $F_0=-Tr\ln \mathcal{G}_0^{-1}/\beta+\mu_{m_F}N_{m_F}$ is the bare free energy of fermions with $\mathcal{G}_0^{-1}=\partial_\tau+\hat{H}_0-\mu_{m_F}$, and the susceptibility $\chi_{m_F}$ is given by
\begin{equation}
    \chi_{m_F}=\int_{\mathrm{BZ}} d^2k  \sum_{ij} n(E_{\mathbf{k}}^{(i)}) \frac{ \big|\langle u_{\mathbf{k}}^{(j)} | \eta(\mathbf{r})| u_{\mathbf{k}}^{(i)} \rangle \big |^2}{E_{\mathbf{k}}^{(j)} - E_{\mathbf{k}}^{(i)} }.
\end{equation}
$E_\mathbf{k}^{(i)}$ and  $|u_\mathbf{k}^{(i)}\rangle$ are the eigenvalues and eigenstates of $\hat{H}_0$ in the $i$th band, respectively. Here, the Fermi-Dirac distribution is $n(E_\mathbf{k}^{(i)})=1/\textrm{exp}[\beta(E_\mathbf{k}^{(i)}-\mu_{m_F})+1]$, which becomes the step function $\Theta(E^{(i)}_\mathbf{k}-\mu_{m_F})$ when the system is at zero temperature. $\mu_{m_F}$ characterizes the Fermi energy and is implicitly dependent on the filling fraction $\nu_{m_F} \equiv N_{m_F}/N_l$, where $N_l$ is the total number of quantum states.

\textit{Phase transition conditions}\textemdash The threshold for the superradiant phase transition is identified in Eq.~(\ref{eq:GrandPotential}) as the point where the coefficient of $\Theta_{m_F}^2$ changes sign~\cite{Nagy2008, Nagy2010}. Therefore, the phase boundary occurs at $\lambda_{m_F}^2=-\tilde{\Delta}_c/4\chi_{m_F}$, and by solving this equation, the critical relative polarization angle can be obtained
\begin{equation}
    \varphi^{\mathrm{crit}}=\arcsin \sqrt{\frac{-\tilde{\Delta}_c/4\chi_{m_F}-1}{ (\alpha_v m_F/2\alpha_s F)^2-1} }.
    \label{eq:crit_condition}
\end{equation}
The value $\varphi^{\mathrm{crit}}$ represents the maximum relative polarization angle for the transition from superradiance to normal phase. To ensure the existence of a finite $\varphi^{\mathrm{crit}}$, $\chi_{m_F}>-\tilde{\Delta}_c/4$ is required. Physically, the susceptibility $\chi_{m_F}$ characterizes the tendency of the normal system toward superradiance. While $\lambda_{m_F}$ affects the superradiance through the adjustment of $\varphi$.

\begin{figure}[t]
   \includegraphics[width=1\linewidth]{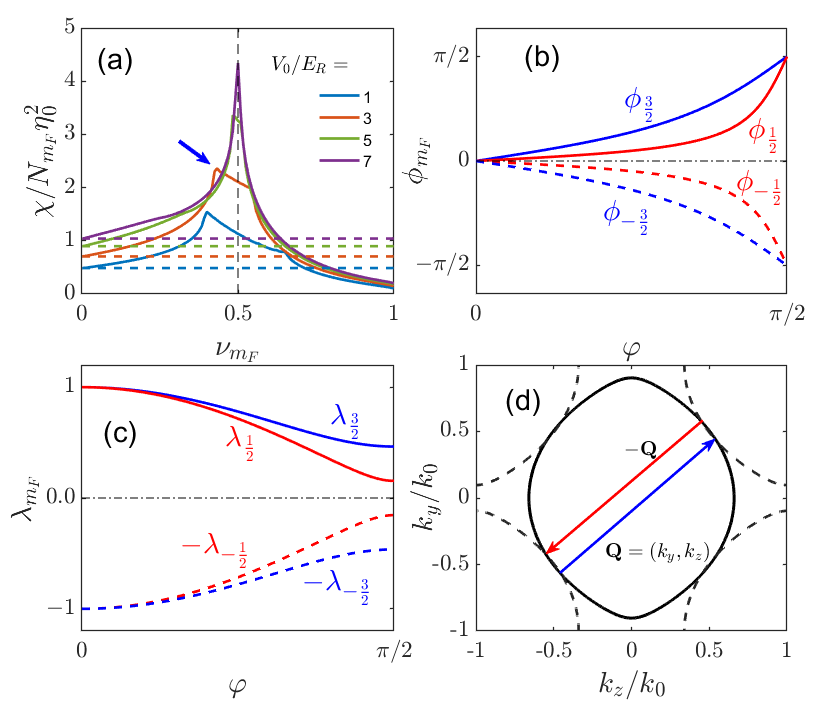}
   \caption{(a) Dependence of the rescaled susceptibility $\chi_{m_F}/N_{m_F}\eta_0^2$  on the filling fraction $\nu_{m_F}$ for several pump depths $\mathrm{V}_0/E_R$. Solid and dashed curves correspond to the noninteracting Fermi and Bose cases, respectively. The inflection points observed in the solid lines are signatures of FS nesting. The blue arrow, located at $\mathrm{V}_0=3E_R$ and $\nu_{m_F}\approx0.43$, denotes the example presented in (d). (b) and (c) Parameters $\phi_{m_F}$ and $\lambda_{m_F}$  plotted as functions of the relative polarization angle $\varphi$. (d)  Nesting of the FS for the Hamiltonian $\hat{H}_0$ before the superradiant transition occurs. Black solid line indicates the original FS, while light blue dashed curves show the FS is shifted by $\mathbf{Q}=\pm\mathbf{k}_p\pm\mathbf{k}_c$. Colored arrows denote the nesting wave vectors $\mathbf{Q}$. }
   \label{fig2}
\end{figure}

\textit{Parameters' influence}\textemdash Fig.~\hyperref[fig2]{2(a)} presents the rescaled susceptibility $\chi_{m_F}/N_{m_F}\eta_0^2$ versus the filling factor $\nu_{m_F}$ and the pump lattice depth $\mathrm{V}_0$, which is independent of the Zeeman states $m_F$. Horizontal dashed lines mark the noninteracting boson case at the same $\mathrm{V}_0$. Data (consistent with Ref.~\cite{ChenYu2014}) show that $\chi_{m_F}$ for the Fermi gas has a strong density dependence and it is in stark contrast to bosonic counterpart; The FS nesting is demonstrated by these peaks of solid lines, and $\chi_{m_F}/N_{m_F}\eta_0^2$ always displays two peaks around $\nu_{m_F}\approx 0.5$. With increasing $\mathrm{V}_0$, fermions cross over from delocalized to localized behavior, and the positions of the peaks move toward $\nu=0.5$. As shown in Fig. 2(d),  we plot the FS of $\hat{H}_0$ at the left peak of $\mathrm{V}_0=3E_R$ (the blue arrow). The part of the FS is well nested with the nesting wave vector $\mathbf{Q}=\pm(\mathbf{k}_c\pm\mathbf{k}_p$).

The inhibitory effect of  $\varphi$ on $\lambda_{m_F}$ is evident in Fig.~\hyperref[fig2]{2(c)}: its value monotonically decreases with increasing $\varphi$. The significance of $m_F$ is manifested in the  evolution of $\lambda_{m_F}$. For a fixed $\varphi$, the reverse $m_F$ state has the same $\lambda_{m_F}$ and a larger $|m_F|$ has a larger $\lambda_{m_F}$. Moreover, $\phi_{m_F}$---also encoding both $\varphi$ and $m_F$---locks the cavity photon phase, see Eq.~(\ref{eq:ave_pho_num}). The sign of $m_F$ determines that of $\phi_{m_F}$, and $|\phi_{m_F}|$ monotonically increases with increasing $\varphi$ and a larger $|m_F|$ corresponds to a larger $\phi_{m_F}$, see Fig.~\hyperref[fig2]{2(b)}.

\begin{figure}[t]
   \includegraphics[width=1\linewidth]{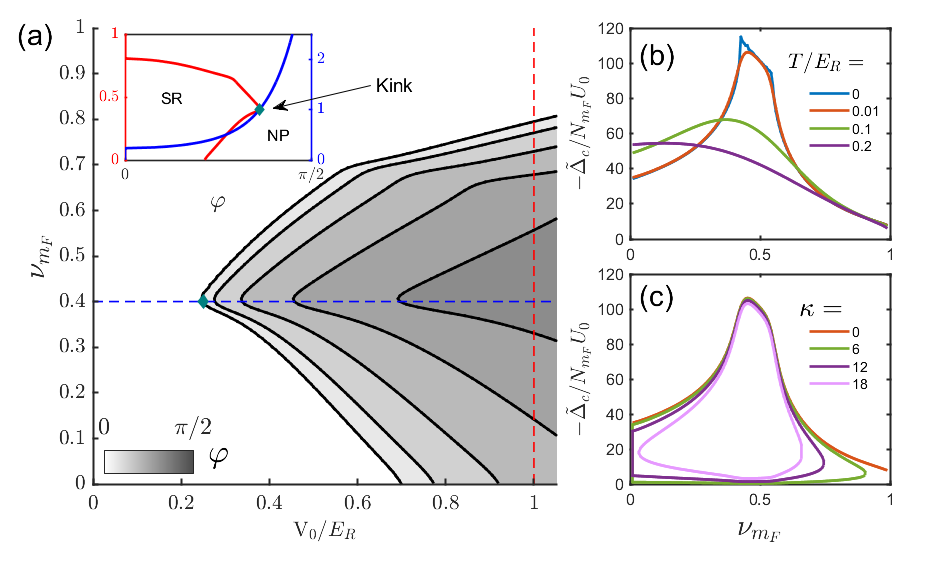}
   \caption{(a) Critical relative polarization angle $\varphi^{\text{crit}}$ as a function of pump lattice depth $V_0/E_R$ and filling fraction $\nu_{m_F}$. Darker shading corresponds to $\varphi^{\text{crit}}$ approaching $\pi/2$; white regions indicate that the superradiant threshold is not reached. Inset: phase diagrams of $\varphi$ and (red) $\nu_{m_F}$ at fixed $\mathrm{V}_{0}=1E_R$ and (blue) $V_0/E_R$ at fixed $\nu_{m_F}=0.40$. The paths are marked by the dashed lines of the same colors in (a). NP and SR denote the normal and superradiant phases, respectively. (b) Evolution with temperature $T$. (c) Impact of cavity losses $\kappa$ at $T=0.01E_R$.}
   \label{fig3}
\end{figure}

\textit{Phase diagrams}\textemdash We have plotted the phase diagrams in Fig.~\ref{fig3} for fixed $m_F=3/2$. The sign of $m_F$ is unimportant and $m_F=1/2$ merely expands the value of  $\varphi^{\textrm{crit}}$; see Eq.~(\ref{eq:crit_condition}). In Figs.~\hyperref[fig3]{3(a)}: for $\varphi^{\textrm{crit}}=0$, the system is in the normal phase due to $\chi_{m_F}<-\tilde{\Delta}_c/4$; for $\varphi^{\textrm{crit}}\neq0$, the system is in the superradiant phase when $\varphi \le \varphi^{\textrm{crit}}$. $\varphi^{\textrm{crit}}=\pi/2$ indicates that for larger $\mathrm{V}_0$ and finite $\nu_{3/2}$, the system is unable to exit superradiance through the suppression arising from $\lambda_{3/2}$. Thus, we find that a minimum $\mathrm{V}_0$ is required (green rhombus). Meanwhile, the critical curve of $\varphi$ (solid lines) suggests the $\chi_{m_F}$ is of strong density dependence. The inset displays the $\varphi$ phase diagram along the $\nu_{3/2}=0.40$ (blue line) and the $\mathrm{V}_0=1E_R$ (red line). The kink represents the value $\varphi^{\textrm{crit}}\approx 0.36\pi$ when $\nu_{3/2}=0.40$ and $\mathrm{V}_0=1E_R$.

Effective heating in such fermion-cavity systems is unavoidable, and the associated broadening of the FS markedly reduces susceptibility $\chi_{m_F}$. Fig.~\hyperref[fig3]{3(b)} shows that raising the temperature flattens the nesting peak and lowers the value $\chi_{3/2}$. Importantly, the nesting peaks at $\nu_{3/2}=0.5$ vanish for temperatures above $T=0.2E_R$.

For non-negligible cavity losses $\kappa$, the phase diagram curve is significantly altered~\cite{Keeling2010, NieXiaotian2023}, and nontrivial phenomenon as limit cycles or chaos may occur~\cite{Piazza2015, Chiacchio2019}. The phase diagram of such an open system is governed by determining the stable attractors of the dynamics. However, the extrema of free energy $F$ in the limit $\kappa \to 0$  coincides with the dynamical stationary point, so the system inherits key features
from its equilibrium counterpart~\cite{Baumann2010, Baumann2011}. When $\kappa \neq 0$, the hard condition for the existence of $\varphi^{crit}$ is rewritten as $\tilde{\Delta}_c=-2\chi_{m_F}\pm\sqrt{4\chi_{m_F}^2-\kappa^2}$. As shown in Figs.~\hyperref[fig3]{3(d)}, the existence of $\kappa$ shrinks the boundary of
stability, yet the peaks of FS nesting survives over a range of $\kappa$.

%\textcolor{Rpcorrect}{For non-negligible cavity losses $\kappa$, the phase diagram curve is significantly altered, and nontrivial phenomenon as limit cycles or chaos may occur. Unlike equilibrium systems, the phase diagram of such an open system is governed by  determining the stable attractors of the dynamics rather than minimizing the free energy. In the limit $\kappa \to 0$, however, the extrema of $F$ coincides with the dynamical stationary point, so the system preserves the key characteristics of the corresponding equilibrium state.

\begin{figure*}[t]
   \includegraphics[width=1\linewidth]{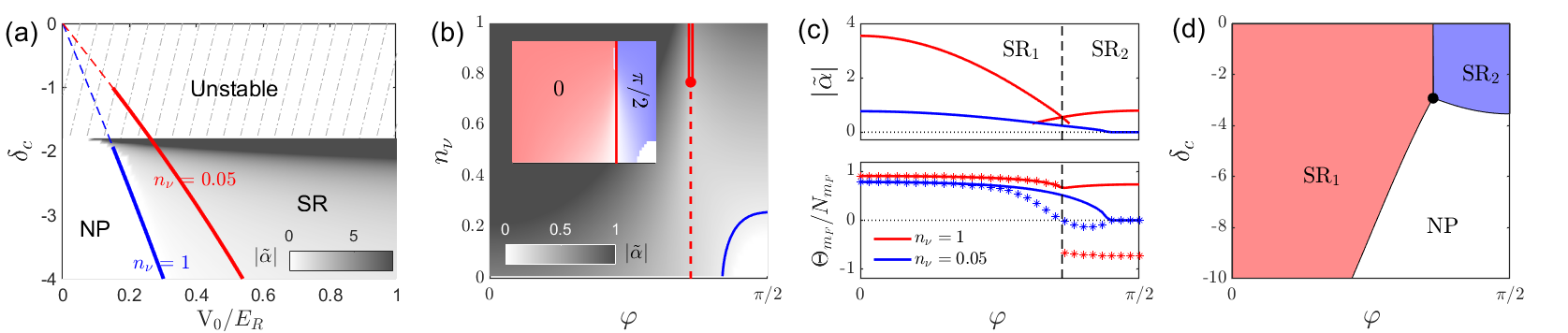}
   \caption{Equilibrium phase diagrams are presented for (a)  $n_\nu=1$ at $\varphi=0$ and (b) the $\varphi$-$n_\nu$ plane with $V_0/E_R=1$ and $\tilde{\delta}_c/E_R=-2.7$. (a) The blue and red curves show second-order boundaries separating the normal (NP) and superradiant (SR) phases for $n_\nu=1$ and $n_\nu=0.05$, respectively. (b) The blue line is the NP-SR boundary. The vertical red line indicates the component-separation boundary~\cite{Ali2022} at $\varphi^{\mathrm{crit}}=\mathrm{arctan}(2\alpha_s/\alpha_v)$. The separation is second order (dashed) for $n_\nu < 0.73$ and first order (double solid) for $n_\nu > 0.77$.  The inset shows the photon phase near 0 ($\pi/2$) for $\varphi < \varphi^{\mathrm{crit}}$ ($\varphi > \varphi^{\mathrm{crit}}$). (c) Instances of (b) at $n_\nu=1$ and $n_\nu=0.05$.  The local minimum of the average photon number $|\tilde{\alpha}|$ (top panel) and the global minimum of the density order parameter $\Theta_{m_F}/N_{m_F}$ (bottom panel) are plotted versus $\varphi$. The black dotted line marks the separation threshold. In the top panel, the red curve ($n_\nu=1$) shows two local minima near threshold, characteristic of a first-order transition, while the blue curve ($n_\nu=0.05$) shows a single minimum, characteristic of a second-order transition. In the bottom panel, the solid line and asterisk correspond to $m_F$ and $-m_F$, respectively. The sign of $\Theta_{m_F}/N_{m_F}$ indicates particle location on odd ($+$) and even ($-$) lattice sites in a checkerboard pattern within the SR phase~\cite{Baumann2010}. At threshold, opposite signs for the two components reveal real-space phase separation. (d) shows the phase diagram in the  $\varphi$-$\tilde{\delta}_c$ plane for $V_0/E_R=1$ and $n_\nu=1$. The black dot indicates the tricritical point.}
   \label{fig4}
\end{figure*}

\textit{Two-component system with reverse $m_F$}\textemdash   Next, we consider a Fermi gas in an arbitrary mixture of the $m_F=3/2$ and $m_F=-3/2$. Compared to Eq.~(\ref{eq:single_component_H}) and Eq.~(\ref{eq:ave_pho_num}), the Hamiltonian and the average photon number for this situation both acquire an additional summation over spin states $\sum_{m_F=\pm3/2}$ preceding the atomic terms. In this case of particle number imbalance, characterized by the ratio $n_{\nu}=\nu_{-3/2} /\nu_{3/2}$, we obtain the phase diagram in Fig.~\ref{fig4} by minimizing $f$. The relationship between free energy and grand potential satisfies the Legendre transformation~\cite{Reichl2016}, and the Helmholz free energy is written as
\begin{equation}
\begin{aligned}
    f=&-\beta^{-1} \int_{\mathrm{BZ}} d^2k \sum_{m_F=\pm\frac{3}{2}}\Big( \sum_{i}\ln[1+e^{-\beta(\epsilon_{\mathbf{k},m_F}^{(i)}-\mu_{m_F})}]\\
&+\mu_{m_F}\nu_{m_F} \Big)-\delta_c|\tilde{\alpha}|^2.
\end{aligned}
\end{equation}
The rescaled cavity field and cavity-pump detuning are given by $\tilde{\alpha}^2=\mathrm{U}_0\alpha^2/4$  and $\delta_c=4\Delta_c/\mathrm{U}_0 N_l$, respectively. Here, $\epsilon_{\mathbf{k},m_F}$ are the eigenvalues of the component in Zeeman state $m_F$, found by diagonalizing the atomic part of $\sum_{m_F}\hat{H}_{\mathrm{eff}}$. We work at a low nonzero temperature to comply with effective heating, $T=0.01E_R$. To simplify, we set $\nu_{3/2}=0.45$---a value near the FS nesting peak---which yields a large $\chi_{3/2}$ to facilitate superradiance.

\textit{Steady-state analysis}\textemdash See Fig.~\hyperref[fig4]{4(c)};  in the superradiant phase the signs of $\Theta_{3/2}$ and $\Theta_{-3/2}$ are opposite for all $\varphi\approx 0.36\pi$. This behavior indicates real-space separation of the fermion components, accompanied by a cavity photon phase shift from nearly $0$ (phase $\mathrm{SR}_1$) to $\pi/2$ (phase $\mathrm{SR}_2$); see inset of Fig.~\hyperref[fig4]{4(b)}. Furthermore, the transition changes from second-order to first-order as $n_\nu$ increases,  and this threshold is located at $n_\nu=0.73$. This can be understood through the equality of the free energy $F(\Theta_{3/2},\Theta_{-3/2})=F(\Theta_{3/2},-\Theta_{-3/2})$, which gives a critical angle $\varphi^{\mathrm{crit}}=\mathrm{arctan}(2\alpha_s/\alpha_v)$ for phase separation~\cite{supp}. For $\varphi=0$, the system reduces to a spin‑independent two‑component Fermi gas. The transition condition is $\tilde{\delta}_c=-4\sum_{m_F}\chi_{m_F}$ (see Ref.~\cite{supp}), and the unstable region is $\tilde{\delta}_c<-2\sum_{m_F}\nu_{m_F}$. Compared with the single‑component case~\cite{Keeling2014}, these results are shown in Fig.~\hyperref[fig4]{4(a)}.

The threshold for the normal-superradiant transition is set by the rescaled cavity-pump detuning $\tilde{\delta}_c$ and the susceptibilities $\chi_{\pm3/2}$; the relative polarization angle $\varphi$ tunes the superradiance of the system and determines the phase separation of the fermion components. Hence, fixing $\varphi=\mathrm{arctan}(2\alpha_s/\alpha_v)$ and selecting appropriate values of $\tilde{\delta}_c$ and $\chi_{\pm3/2}$, one obtains tricritical points connecting the normal, $\mathrm{SR}_1$, and $\mathrm{SR}_2$. For simplicity, we present the phase diagram for $n_\nu=1$ in Fig.~\hyperref[fig4]{4(d)}. In this case, two components equally occupy on one site with zero magnetization for $\varphi<0.36\pi$, and for $\varphi>0.36\pi$ they display alternating distribution characteristic of antiferromagnet. Therefore, the critical thresholds between zero magnetization and antiferromagnet phases can be obtain in the mean-field approach~\cite{Landini2018}: For the former, the critical effective pump-cavity detuning is found algebraically $\delta_c^{\text{crit}}=-8\chi\cos^2(\phi)$; for the latter, the critical effective pump-cavity detuning is instead  $\delta_c^{\text{crit}}=-8\chi\sin^2(\phi)$, where we denote $\chi_{\pm3/2}=\chi$~\cite{supp}. Retraced to the case of BEC~\cite{Landini2018, Chiacchio2019}, the same nature will remain. Yet, the tricritical point (the threshold of phase separation) is shifted for finite dissipation, depending on the number imbalance between two components due to $\chi_{m_F} \propto  N_{m_F}$~\cite{supp}.

\textit{Conclusions and outlook.}\textemdash This work clarifies the role of the relative polarization angle $\varphi$ in a spin-dependent degenerate Fermi gas confined in an optical cavity, where the cavity and pump polarizations are misaligned.  The atom-light interaction, governed by $\varphi$, mediates the balance between scalar and vectorial contributions, thereby enhancing superradiance for $\alpha_v/\alpha_s>1$ and suppressing it for $\alpha_v/\alpha_s<1$. We also systematically characterize the effects of the filling factor $\nu_{m_F}$, cavity detuning $\tilde{\Delta}_c$, pump lattice depth $\mathrm{V}_0$, finite temperature $T$, and cavity dissipation $\kappa$ on the phase diagram.  In a binary Fermi mixture with opposite spins, real-space phase separation invariably occurs at $\varphi = \text{arctan}(2\alpha_s/\alpha_v)$. When the particle numbers are imbalanced, the transition crosses over from second order to first order as the ratio $n_\nu$ increases. These findings are equally applicable to the bosonic case. While the principle of free-energy minimization can detect the onset of superradiance~\cite{Farokh2021,RMP2021}, it is fundamentally inadequate for describing the long-time steady state of this system, as the interplay of cavity losses and Pauli blocking can lead to nonthermal steady states~\cite{PIAZZA14A,Zwettler2025}. The dissipation-driven nonequilibrium dynamics, which determines the behavior of the unstable regions through long-term evolution, has not been fully explored and will constitute the focus of our subsequent research. Though mean-field theory is usually a good approximation in long-range interacting systems~\cite{RMP2023}, quantum fluctuations can modify the critical exponents~\cite{RMP2025}, and drive the system to a novel dynamical phase~\cite{DIEHL10} or a bistability~\cite{CATALIN25}. Extending our model to include additional interactions represents a natural step toward investigating  stoner ferromagnetism~\cite{SIMON09,KETTERLE12}, strong correlations~\cite{Landig2016, Helson2023}, BCS superconductors~\cite{Young2024}, non-Hermitian phenomena~\cite{David2026} and related topics.
%and can be observed using a balanced heterodyne detection scheme for the phase and amplitude of the intracavity field~\cite{DONNER2025} and an electron‑enhanced charge‑coupled device for the spatial distribution of the atomic cloud~\cite{ZhangXiaotian2021}.
\begin{acknowledgments}
\emph{Acknowledgments}---%
This work is supported by the National Natural Science Foundation of China under Grants No.~12174055 and No.~11674058, and by the Natural Science Foundation of Fujian Province under Grant No.~2025J01658.

%\emph{Data availability}---%
%The data that support the findings are available from the authors upon reasonable request.
\end{acknowledgments}
%\bibliography{RefQin}

\end{document}